\documentclass[12pt]{article}
\usepackage{epsfig}
\usepackage{latexsym}
\begin{document}
\title{Decoupling of the longitudinal polarization of the vector field\\
in the massless Higgs-Kibble model.}
\author{ A.A.Slavnov \thanks{E-mail:$~~$ slavnov@mi.ras.ru}
\\Steklov Mathematical Institute, Russian Academy
of Sciences\\ Gubkina st.8, GSP-1,119991, Moscow} \maketitle

\begin{abstract}
It is shown that the three dimensionally longitudinal component of
the vector field decouples in the massless limit of nonabelian Higgs
model.
\end{abstract}

\section{Introduction.}

It was known long ago that the massless limit of the global $U(1)$
invariant theory of massive neutral vector field interacting with
the matter fields reproduces the results of quantum electrodynamics
(QED). In the limit $\mu\rightarrow 0$ the three dimensionally
longitudinal component of the vector field decouples of all other
excitations and we are left with two threedimensionally transversal
polarizations of the photon interacting with matter fields.

In the nonabelian case the situation is different. In the limit
$\mu\rightarrow 0$ the massive Yang-Mills model the three
dimensionally longitudinal polarization of the vector field does not
decouple  \cite{SF}. At one loop it gives a finite contribution and
at higher loops in the limit $\mu\rightarrow 0$ we have pole
singularities c. It is natural to suggest that the massless limit
can be achieved in the Higgs model, leading to the massless
Yang-Mills field interacting with the massless Higgs meson.

Recently in the paper \cite{RF} it was stated that in the limit
$\mu\rightarrow 0$ of the nonabelian Higgs model the longitudinal
component of the vector field does not decouple, but undergoes the
metamorphosis to the massless scalar fields, corresponding to the
Goldstone bosons of the Higgs model.

In this paper we argue that the hypothesis about decoupling of the
longitudinal component of the vector field in the massless limit of
the Higgs-Kibble  model is correct. The statement of the paper
\cite{RF} corresponds to another massless limit: massless Yang-Mills
field interacting with the complex scalar doublet with real nonzero
mass, when this mass  vanishes. Although if the mass of the
Yang-Mills field is less than the accuracy of measurement both these
scenarios correspond in the hypothetical limit $\mu\rightarrow 0$ to
the massless Yang-Mills theory, the number of physical degrees of
freedom is different and these two scenarios may be distinguished
experimentally.

\section{Decoupling of the longitudinal component of the vector field in
the Higgs-Kibble model in the limit $\mu=0$.}

 The Higgs-Kibble model is gauge invariant and may be considered in different
 gauges. In the unitary gauge $\varphi^a=0$, where
 \begin{equation}
\varphi=(\frac{i\varphi_1+\varphi_2}{\sqrt{2}}, \quad
\frac{\sigma-i\varphi_3}{\sqrt{2}})
 \label{1}
\end{equation}
the Lagrangian has a form
\begin{eqnarray}
L=-\frac{1}{4}F_{\mu\nu}^a F_{\mu\nu}^a
+\frac{m_1^2}{2}A_{\mu}^aA_{\mu}^a
+\frac{1}{2}\partial_{\mu}\sigma \partial_{\mu}\sigma-\frac{m_2^2}{2}\sigma^2\nonumber\\
+\frac{m_1g}{2}\sigma A_{\mu}^2+\frac{g^2}{8}\sigma^2
A_{\mu}^2-\frac{gm_2^2}{4m_1}\sigma^3-\frac{g^2m_2^2}{32m_1^2}\sigma^4\nonumber\\
m_1=\frac{\mu g}{\sqrt{2}}; \quad m_2=2\lambda \mu
 \label{2}
\end{eqnarray}
In this gauge the spectrum of the model is obvious: nine massive
excitations corresponding to massive vector field $A_{\mu}^a$, and
one scalar field $\sigma$. Goldstone bosons $\varphi^a$ and
dynamical ghost fields $c,\bar{c}$ are absent.

It is known that although this gauge is not manifestly renormalizable,
for any invariant regularization all nonrenormalizable divergencies
present in observable gauge invariant amplitudes cancel and to calculate
a renormalized scattering amplitude it is sufficient to redefine the
parameters entering the Lagrangian (\ref{2}): masses, charge and wave
function normalization of the fields.

We shall be interested in the amplitudes for the massless fields, when
the parameter $\mu=0$. These amplitudes are infected by the infrared
divergencies, but we shall ignore these logarithmic divergencies,
assuming that $\mu$ is small but finite. These divergencies are absent in
the gauge invariant Green functions, whose value after renormalization
also does not depend on a gauge.

To avoid a necessity to look for cancelation of nonrenormalizable
divergencies in observable amplitudes we shall use not unitary, but the
Lorentz gauge $\partial_{\mu}A_{\mu}^a=0$. The corresponding effective
Lagrangian looks as follows
\begin{eqnarray}
L_{ef}=-\frac{1}{4}F_{\mu\nu}^a F_{\mu\nu}^a+(D_{\mu}\varphi-
\hat{\mu})^*(D_{\mu}\varphi- \hat{\mu})-
\kappa[(\varphi^*-\hat{\mu})(\varphi-\hat{\mu})-\frac{\mu^2}{g^2}]^2\nonumber\\
+\lambda^a\partial_{\mu}A_{\mu}^a+i\bar{c}^a\partial_{\mu}D_{\mu}c^a;
\quad \hat{\mu}=(0,\frac{\mu \sqrt{2}}{g}). \label{3}
\end{eqnarray}
Here $\bar{c}^a$ and $c^a$ are Faddeev-Popov ghosts described by the
anticommuting scalar fields.

The spectrum of the theory in this gauge includes apart from physical
quanta also unphysical component of the vector field, Faddeev-Popov
ghosts and Goldstone bosons $\varphi^a$.

The Lagrangian (\ref{3}) is invariant with respect to the BRST
transformations
\begin{eqnarray}
\delta A_{\mu}^a=(D_{\mu}c)^a \nonumber\\
\delta \varphi^a=\mu
c^a+\frac{g}{2}\varepsilon^{abd}\varphi^bc^d+\frac{g}{2}\sigma c^a
\nonumber\\
\delta \bar{c}^a=\lambda^a \nonumber\\
\delta c^a=g\varepsilon^{abd}c^bc^d \nonumber\\
\delta \sigma=-\frac{g}{2}\varphi^ac^a\nonumber\\
\delta \lambda^a=0. \label{4}
\end{eqnarray}
By the Noether theorem this invariance generates the conserved charge
$Q$, and physical asymptotic states are selected by the condition
\begin{equation}
Q^0|\varphi>_{as}=0 \label{5}
\end{equation}
where $Q^0$ is the asymptotic BRST charge.

For any $\mu$ different from zero the invariance of the Lagrangian
(\ref{3}) with respect to the transformations (\ref{4}) leads to
decoupling of the nine physical vector field excitations and the
excitation corresponding to the scalar field $\sigma$ from the
excitations corresponding to unphysical components of the vector field,
Faddeev-Popov ghosts and the Goldstone bosons $\varphi^a$ \cite{KO}.

The field $A_{\mu}^a$ may be decomposed into the components with
different polarizations
\begin{equation}
A^a_i(\textbf{k})=e^1_ia^1_a+e^2_ia^a_2+\frac{k_ik_0}{|k|m_1}a^a_3; \quad
A^a_0(\textbf{k})=\frac{|k|}{m_1}a_3^a \label{6}
\end{equation}
In the eq.(\ref{6}) the vectors $e_1$ and $e_2$ are the unit vectors
orthogonal to each other and to the momentum $\textbf{k}$.

Having in mind that the propagator of the vector field in the
transversal gauge looks as follows
\begin{equation}
D_{ab}^{\mu\nu}=\delta_{ab}\frac{g^{\mu\nu}-k^\mu k^\nu
k^{-2}}{k^2-m^2} \label{6a}
\end{equation}
we see that the amplitude, describing the transition from the
transversally polarized state to another transversally polarized
state is finite in the limit $\mu=0$. On the other hand the matrix
elements between states including the longitudinal components of the
vector field in general are singular in the limit $\mu\rightarrow
0$. Nevertheless, as we shall show, in the limit $\mu\rightarrow 0$
the longitudinal states decouple and the scattering matrix is
unitary in the space including only three dimensionally transversal
polarizations and the massless Higgs scalar.

 Let us consider the forward scattering between states which do not
include longitudinal states. As follows from the discussion above,
we may define the unitary scattering matrix by using the projectors
to the states including only the three polarizations of the massive
vector field and the massive Higgs meson: $\tilde{S}=PSP$. The
matrix element describing the forward scattering of the transversal
polarizations of the vector field and the Higgs mesons has a
definite limit when $\mu\rightarrow 0$. By the optical theorem the
imaginary part of the amplitude of the forward scattering is
proportional to the total cross section of the process. That means
that the sum
\begin{equation}
\sum_l|<n|\tilde{S}|l>|^2 \sim Im<n|\tilde{S}|n> \label{7}
\end{equation}
where the vectors $|n>$ do not include the longitudinal
polarizations, and the vectors $|l>$ span the complete space where
the scattering matrix acts. The r.h.s. of the equation (\ref{7}) has
a limit when $\mu\rightarrow 0$ and this limit coincides with the
imaginary part of the corresponding amplitude in the massless
theory. This theory is known to be unitary in the space including
only transversal polarizations of the vector field and massless
Higgs meson. As all the vectors including only physical components
of the vector field and the Higgs field have positive definite norms
we conclude that the amplitudes $<n|\tilde{S}|m>=0$ if the vector
$|m>$ contains at least one longitudinally polarized quant. That
means that as it happens in the case of neutral vector meson
longitudinal polarizations of the massive Yang-Mills field decouple
in the limit $\mu\rightarrow 0$, and the resulting theory describes
in this limit the massless Yang-Mills field interacting with the
massless Higgs bosons.

It is worth to mention that the same result may be obtained starting
from the unitary gauge. One should remember however that in the
unitary gauge the variables $A_0^a$ are not dynamical and should be
excluded by solving the constraint equation before the quantization.

Namely, the equation of motion for $A_0$ does not contain time derivative
of canonical variables
\begin{equation}
(m_1+\frac{g\sigma}{2})^2A_0=D_kF_{0k}; \quad F_{0k}=P_k. \label{7a}
\end{equation}
Solving this equation and substituting the solution to the
Lagrangian (\ref{2}) we get the theory described by the
nondegenerate Lagrangian, which may be quantized in a usual way. The
equation (\ref{7a}) is singular in the limit $\mu\rightarrow 0$ and
one should look for the compensation of the singularities generated
by the solution of the eq.(\ref{7a}) and the singularities generated
by the decomposition of the field $A_\mu$ to the states with
different polarizations. Using the Lorentz gauge we avoid this
complicated procedure.

 A massless Yang-Mills field interacting with
the massless scalar may be also described by the other model. Let us
consider the theory of the Yang-Mills field with zero mass
interacting with the scalar complex doublet with a real mass. This
theory is also BRST invariant, but in this case the scalar fields
$\varphi^a$ do not contribute to the asymptotic BRST charge, which
has a form
\begin{equation}
\hat{Q}_B^0=\int d^3k[(a_0^++a_3^+)c^-+c^+(a_0^-+a_3^-)] \label{8}
\end{equation}
As usual one may introduce the operator
\begin{equation}
\hat{K}=\int d^3k[(a_0^+-a_3^+)\bar{c}^-+\bar{c}^+(a_0^--a_3^-)]
\label{9}
\end{equation}
The number operator for the unphysical states may be presented as
anticommutator of the operators $\hat{Q}_0$ and $\hat{K}$
\begin{equation}
\hat{N}_{unph}=[\hat{Q}_0,\hat{K}]_+ \label{10}
\end{equation}
If the number of unphysical states is different from zero
\begin{equation}
\frac{\hat{N}}{N}|\varphi>_{as}=\hat{Q}_0|\chi> \label{11}
\end{equation}
where
\begin{equation}
|\chi>=\hat{K}|\varphi>_{as} \label{12}
\end{equation}
Therefore any state annihilated by $\hat{Q}_0$  may be presented in the
form
\begin{equation}
|\varphi>=|\varphi>_{ph}+Q_0|\chi> \label{20}
\end{equation}
where $|\varphi>_{ph}$ contains only three dimensionally transversal
components of $A_{\mu}^a$ and the excitations corresponding to the
complex doublet of the scalar mesons.

In the limit when the mass of the scalar mesons vanishes we have
precisely the particle content described in the paper \cite{RF}. If the
parameter $\mu$ is small, smaller than the accuracy of measurement, one
cannot distinguish experimentally the transversal polarizations of the
Yang-Mills field in these two scenarios, but the number of degrees of
freedom is different. In the first case in the limit $\mu\rightarrow 0$
the longitudinal quanta decouple, whereas in the second case the charged
scalar bosons interact with the massless Yang-Mills field even in the
limit $\mu\rightarrow 0$.

\section{Conclusion.}

In this paper we showed that the three dimensionally longitudinal
component of the vector field decouples of all other excitations in
the massless limit of the nonabelian Higgs-Kibble model. This is in
complete analogy with the Abelian case. The massless Yang-Mills
theory interacting with massless scalar fields may be also obtained
starting from the Yang-Mills field interacting with the complex
scalar doublet of real mass, when this mass tends to zero. In this
case the number of observable degrees of freedom is bigger: the
charged scalar mesons of zero mass are also present.

{\bf Acknowledgements.} \\I wish to thank R.Ferrari for numerous
discussions, which resulted in the appearance of this paper. This
paper was supported in part by RFBR under grants 11-01-00296a and
11-01-12037 ofi-m--2011 , by the grant of support of leading
scientific schools NS-4612.2012.1 and by the program "Nonlinear
dynamics". $$ ~
$$
 \end{document}